\documentclass[fleqn,11pt]{wlscirep} 

\usepackage[utf8]{inputenc}
\usepackage[T1]{fontenc}
\usepackage{caption}
\usepackage{subcaption}
\usepackage{amsmath}
\usepackage{amssymb}
\usepackage{float}
\usepackage[export]{adjustbox}
\usepackage{array}

\usepackage{lineno}

\usepackage{setspace}
\usepackage{ntheorem}
\usepackage{nameref,hyperref}

\UseMicrotypeSet[protrusion]{basicmath}
\microtypecontext{spacing=nonfrench}
\microtypesetup{nopatch=footnote}

\title{
Beyond Firms and Industries: Shock Propagation through Establishment- and Product-Level Supply Chains
}

\author[1,2*]{Hiroyasu Inoue}
\author[3,4]{Yasuyuki Todo}
\affil[1]{University of Hyogo, Graduate School of Information Science, Kobe, 6500047, Japan}
\affil[2]{RIKEN, Center for Computational Science, Kobe, 6500047, Japan}
\affil[3]{Waseda University, Graduate School of Economics, Tokyo, 1698050, Japan}
\affil[4]{Research Institute of Economy, Trade and Industry, Tokyo, 1008901, Japan}

\affil[*]{inoue@gsis.u-hyogo.ac.jp}

\begin{abstract}
This paper investigates how the granularity of supply-chain data affects the propagation of economic shocks through production networks. Using newly constructed establishment-level supply chains with product-level information links for Japan, we simulate disruption dynamics under alternative definitions of network nodes and input classifications. We show that defining inputs at the product level generates substantially larger propagation effects than industry-based classifications, indicating that coarse industry measures overstate input substitutability and underestimate systemic vulnerability. While establishment-level networks generally amplify shock propagation relative to firm-level networks, this effect is quantitatively modest, reflecting opposing forces of increased network complexity and greater substitution possibilities. We further demonstrate that incorporating establishment-level geographic information is critical for assessing region-specific shocks, as firm-level networks tend to overstate the impact of shocks originating in major metropolitan areas. Overall, our results highlight the importance of granular information on products, establishments, and geography for accurately evaluating supply-chain resilience and systemic risk.

\vspace{10pt}
\end{abstract}

\begin{document}

\doublespacing

\maketitle

\section{Introduction}

Disruptions to supply chains pose significant risks to the real economy because their effects propagate through networks of production relationships. When an exogenous shock---such as a natural disaster, pandemic, geopolitical conflict, or protectionist policy---reduces the output of a subset of firms, upstream suppliers experience declines in demand, while downstream customers face shortages of intermediate inputs. These disruptions can trigger reductions in production among both suppliers and customers of directly affected firms. \cite{Carvalho2016,Barrot2016} As a result, even a localized or seemingly minor shock may generate substantial indirect effects---often exceeding the direct impact of the initial disturbance \cite{tierney1997business,pelling2002macro}---and give rise to aggregate fluctuations at the macroeconomic level. \cite{bak1993aggregate,gatti2005new,acemoglu2012network}

A large body of earlier research relies on inter-industry analyses based on input--output (IO) tables to study the propagation of shocks through supply chains. \cite{acemoglu2012network,haimes2001leontief,okuyama2004measuring} While these models offer a comprehensive representation of production linkages at the sectoral level, IO tables aggregate heterogeneous firm-level transactions within industries. As a consequence, they abstract from firm heterogeneity and the complex structure of supply chains, limiting their ability to capture cascading failures and nonlinear propagation dynamics. This aggregation has been shown to lead to underestimation of systemic effects \cite{inoue_firm-level_2019,Bossut2024granular} and, in some cases, to misleading conclusions regarding the magnitude and distribution of shock impacts. \cite{diem2024estimating}

To overcome these limitations, recent studies increasingly exploit firm-level supply-chain data made available by advances in data collection and processing. \cite{poledna2018does,inoue_firm-level_2019,diem2022quantifying,pichler_building_2023} Analyses based on firm-level networks consistently find that disruptions propagate more strongly than predicted by industry-level IO models. \cite{inoue2019firm,diem2024estimating} This literature also emphasizes the importance of input substitutability in mitigating propagation effects, highlighting how firms’ ability to replace disrupted suppliers shapes aggregate outcomes. \cite{inoue2019firm,inoue2023trade}

Despite these advances, existing firm-level analyses remain subject to important limitations. First, production and transactions are conducted at the establishment level rather than the firm level, particularly for multi-establishment firms. A supply-chain link observed between two firms does not necessarily imply transactions between all possible pairs of their establishments. Firm-level networks therefore impose an artificial aggregation that may obscure the true structure of production relationships and distort estimates of propagation effects.

Second, firm-level studies typically proxy a firm’s output and inputs using broad industry classifications. In practice, firms often produce multiple products, and different establishments within the same firm may specialize in distinct products. Because input substitution depends on technological and product-level compatibility, coarse industry classifications may substantially overstate substitutability and thereby understate the true extent of shock propagation. \cite{Berthou2024granular}

Third, firm-level data provide limited information on the geographic location of production activities. Because firms often operate establishments in regions different from their headquarters, shocks that are geographically localized---such as earthquakes or regional lockdowns---may be misidentified when only firm-level locations are observed. This limitation may lead to systematic bias in the estimated effects of regional shocks.

To address these limitations, we construct a granular supply-chain network with establishment-level nodes classifying inputs by product, rather than industry using three large-scale datasets for Japan. The first two datasets are the Company Information Database and the Company Linkage Database collected by Tokyo Shoko Research (TSR), covering 1{,}520{,}605 firms and 5{,}860{,}726 supplier--client relationships in 2020. Focusing on manufacturing firms, we combine these data with the Economic Census for Business Activity conducted by the Ministry of Economy, Trade and Industry, which provides detailed product information for 512{,}401 manufacturing establishments in 2021. By integrating firm-level links with establishment-level product information, we infer feasible production recipes and identify the most likely establishment-level supply-chain links. The resulting network comprises 183{,}951 establishments belonging to 157{,}537 firms and 919{,}982 establishment-level links with product information.

We then apply a probabilistic model of supply-chain disruption to these networks. In the model, when a supplier ceases production, a downstream establishment becomes inactive with a probability that decreases in the share of alternative suppliers producing the same input. This framework allows us to explicitly capture the role of input substitutability in shaping propagation dynamics.

Our simulation analysis delivers several key findings. First, propagation effects are substantially larger when inputs are classified at the product level than when they are defined by broad industry categories, indicating that industry-level classifications significantly understate systemic vulnerability. Second, while establishment-level networks generally exhibit stronger propagation than firm-level networks, the magnitude of this difference is modest and reflects opposing forces: increased complexity of networks that amplifies diffusion and greater opportunities for substitution that mitigate it. Third, incorporating establishment-level geographic information is crucial for evaluating region-specific shocks. Firm-level networks tend to overestimate the impact of shocks originating in Tokyo and underestimate those occurring elsewhere, owing to the concentration of corporate headquarters in Tokyo relative to production establishments.

Taken together, these results demonstrate that accurately assessing the propagation of supply-chain disruptions requires fine-grained information on products, establishments, and geography. Analyses based solely on firm-level or industry-level data risk mischaracterizing both the magnitude and spatial distribution of economic disruptions. Our establishment-level, product-specific framework provides a more precise foundation for evaluating systemic risk and economic resilience in modern supply chains.

\section{Data}

\subsection{Primary data}

The primary data source for this study is a set of firm-level datasets collected by Tokyo Shoko Research (TSR), consisting of the Company Information Database and the Company Linkage Database. The Company Information Database provides firm attributes, including company name, address, industry classification, sales, and corporate number. The Company Linkage Database reports domestic supplier--client relationships between firms. We use the 2020 vintage of the TSR data, which is the most recent year available and closely aligns with the timing of the Economic Census described below. The 2020 TSR data cover 1{,}520{,}605 firms and 5{,}860{,}726 trade links, capturing the vast majority of economically relevant firms in Japan, with the exception of micro-enterprises. From this dataset, we later extract firms in the manufacturing sector.

While the TSR data provide comprehensive coverage of firm-level supply-chain relationships, they lack two key dimensions that are central to our analysis: establishment-level information and detailed product-level output data. To address these limitations, we combine the TSR data with the Economic Census for Business Activity (hereafter, the census), conducted every five years by the Ministry of Internal Affairs and Communications and the Ministry of Economy, Trade and Industry of Japan \cite{SBJ2021}. The census covers all establishments across all industries, including micro, small, and medium-sized enterprises, and reports detailed establishment-level information such as location, employment, business activities, sales, and expenditures. Crucially for our purposes, the census identifies the specific products produced by each establishment. Note that an establishment may have multiple output products. However, it does not provide information on trading partners, leaving establishment-level supply-chain links unobserved.

We use the 2021 wave of the census, which includes 5{,}156{,}063 establishments. Using corporate ID numbers, we merge the census with the TSR data, thereby augmenting the TSR firm-level network with information on 1{,}014{,}673 non-headquarter establishments. Although the merged dataset covers the universe of firms and establishments in Japan, we restrict our analysis to the manufacturing sector. This restriction is motivated by both data availability and methodological considerations: identifying product-level production recipes is more tractable in manufacturing, whereas extending the approach to other sectors---particularly wholesale---raises additional challenges in interpreting input--output relationships. The resulting manufacturing sample comprises 157{,}537 firms and 183{,}951 establishments.

\subsection{Constructing product-level recipes}
\label{ch:recipe}

The TSR data identify firm-level trade relationships, while the census provides establishment-level product information but does not report trade links. As a result, there is no direct observation of supply-chain relationships between establishments. Assuming that all establishments of a supplier firm trade with all establishments of a client firm would generate an implausibly dense network and misrepresent actual production relationships, especially for multi-establishment firms producing heterogeneous products. We therefore infer establishment-level connections by identifying plausible links from the full set of potential supplier--client establishment pairs.

The key first step in this process is the construction of product-level production recipes, which allows us to exclude establishments that do not supply relevant inputs. We use three sources of information: firm-level supply-chain relationships from the TSR data, firm--establishment mappings from the census, and establishment-level output products from the census. Each firm operates one or more establishments, and each establishment produces one or more products. As an initial step, we connect all establishments of supplier firms to all establishments of client firms. For each output product ($P^{\text{\scriptsize out}}$) of an establishment, we then identify candidate input products ($P^{\text{\scriptsize in}}$) based on the products produced by the linked supplier establishments. This procedure yields, for each output product, a set of candidate inputs and the number of times ($\mathit{obs.}$) each input is observed across establishments producing that output.

The number of observations $\mathit{obs.}$ ranges from zero to the total number of establishments ($\mathit{est.}$) producing the output product. Because supplier firms may operate multiple establishments producing unrelated products, this process may introduce spurious inputs. To filter out such noise, we impose a 50\% cutoff rule: an input product is included in the recipe for an output product only if it appears in more than half of the establishments producing that output. Formally, the set of inferred recipes, or valid input-output product pairs $(P^{\text{\scriptsize in}}, P^{\text{\scriptsize out}})$, is defined as
\begin{equation}
\{(P^{\text{\scriptsize in}}, P^{\text{\scriptsize out}}) \mid \mathit{obs.} > \mathit{est.}/2\}.
\end{equation}

To illustrate, suppose that {\it aluminum rod} is produced by 10 establishments. If all 10 are linked to suppliers producing {\it aluminum ingot}, while only 3 and 1 are linked to suppliers producing {\it degasser pellets} and {\it titanium ingot}, respectively, then only {\it aluminum ingot} satisfies the cutoff criterion and is included in the recipe for {\it aluminum rod}. The other products are excluded as non-essential inputs.

This approach has limitations. First, inputs that are used by only a subset of establishments---such as quality-enhancing additives---may be excluded despite being technologically relevant. Second, when establishments engage in multi-stage production processes, some intermediate inputs may not be directly observed as supplier outputs. For example, producers of {\it aluminum rod} may purchase {\it alumina} and internally produce {\it aluminum ingot}, which would lead to the exclusion of {\it aluminum ingot} from the inferred recipe despite its technological importance. These limitations are inherent to inference based on observed transactions and are discussed further in the Conclusion.

\subsection{Constructing the establishment- and product-level network}
\label{ch:construct}

Using the inferred product-level recipes, we construct establishment-level supply-chain networks in two steps. First, we eliminate any potential establishment-level link for which none of the supplier establishment’s products appear in the recipe of the client establishment’s output products. This step substantially reduces the number of candidate links but may still generate redundant connections. For example, if a client establishment is linked to a supplier firm with multiple establishments producing identical products, all such establishments may be retained despite only one actual supplier being required.

To further refine the network, we formulate a set cover problem with group and priority constraints. The objective is to select, for each client establishment, the smallest set of supplier establishments whose combined products cover all required inputs. The universal set consists of the input products identified by the recipe, and each subset corresponds to a supplier establishment and its product portfolio. Although the set cover problem is NP-hard, we obtain approximate solutions using a greedy algorithm, which is computationally feasible given the limited size of each instance. Details of the algorithm are provided in SI Section~1.

We impose two additional constraints. First, we introduce group constraints, where each group corresponds to a supplier firm. The algorithm must select at least one establishment from each group, reflecting the existence of an observed firm-level supplier--client relationship. Second, we introduce a priority mechanism that penalizes repeated selection of establishments belonging to the same firm, encouraging diversification across supplier firms when multiple establishments offer identical inputs.

The resulting network represents a parsimonious configuration of establishment-level supply chains that satisfies all inferred input requirements. Because establishment-level supply-chain data are not observed, the network should be interpreted as one plausible realization rather than a definitive mapping. In practice, establishments may maintain redundant supplier relationships for risk management or contractual reasons, which our construction does not capture.

\subsection{Description of firm- and establishment-level supply chains}

In the manufacturing sector, there are 2{,}396 distinct product classifications and 595 four-digit industry classifications under the Japan Standard Industrial Classification (JSIC) \cite{JSIC2013}. Based on firm-level supply-chain relationships, the total number of observed input--output product combinations is 1{,}182{,}274. Applying the cutoff procedure described in Section~\ref{ch:recipe} reduces this set to 43{,}340 inferred product-level recipes. The full set of potential establishment-level links implied by firm-level relationships and establishment mappings amounts to 18{,}872{,}233.

\begin{figure}[tbp]
\centering
\begin{minipage}{0.5\textwidth}
    \centering
    \includegraphics[width=\linewidth]{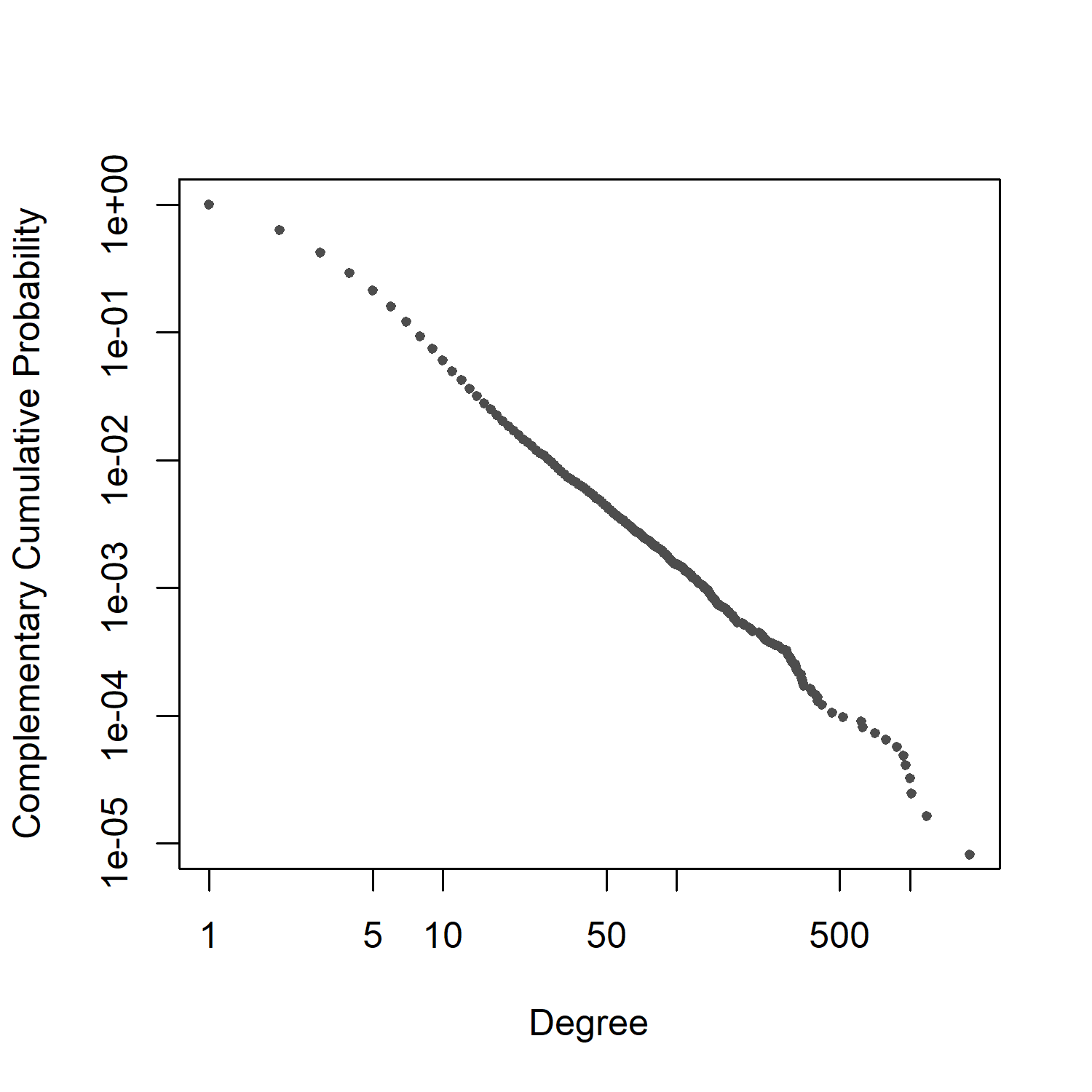}
    \vspace{-2em} 
    \subcaption{Firm level}
\end{minipage}%
\begin{minipage}{0.5\textwidth}
    \centering
    \includegraphics[width=\linewidth]{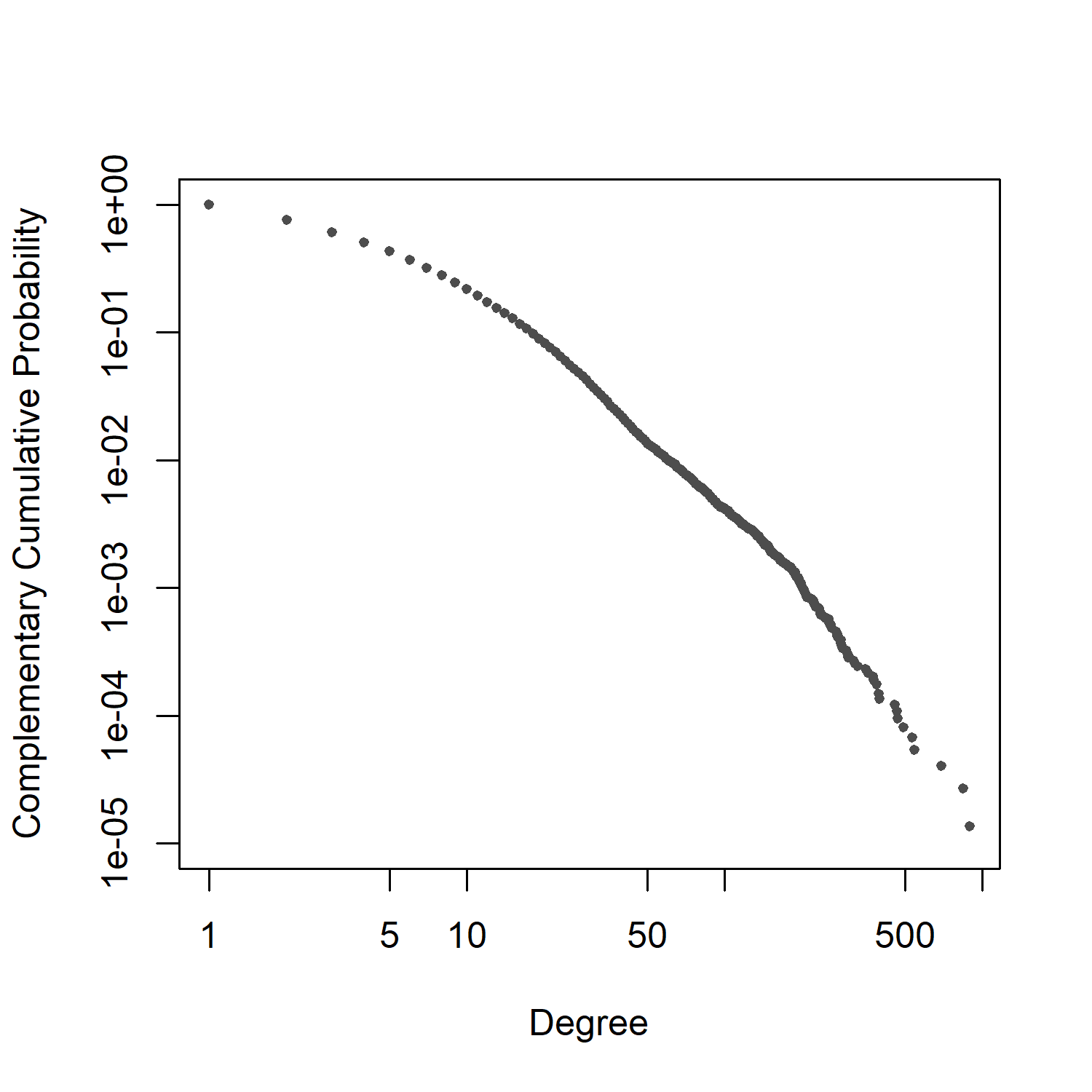}
    \vspace{-2em} 
    \subcaption{Establishment level}
\end{minipage}
\caption{Degree of production networks}
\label{fig:deg}
\end{figure}

After applying the filtering and optimization procedures described above, the establishment-level network consists of 183{,}951 establishments and 919{,}982 links, compared with 157{,}537 firms and 480{,}507 links in the original firm-level network. As shown in Table~\ref{tbl:networkcount}, the median number of links differs only slightly between the firm-level and the establishment-level networks. By contrast, the 90th percentile of the number of links per node is substantially higher in the establishment-level network than in the firm-level network. Specifically, the 90th percentile is 7 in the firm-level network and 17 in the establishment-level network, indicating that the establishment-level network contains a larger number of highly connected nodes. This difference reflects the aggregation of multiple establishment-level links within the same firm into a single firm-level link in the TSR data. Naturally, this contrast becomes even more pronounced when considering the complete set of potential establishment-level links (SI Figure~1). Despite differences in scale, both networks exhibit degree distributions that closely follow power-law patterns (Figure~\ref{fig:deg}), implying short average path lengths and the potential for rapid shock propagation, as suggested in the literature \cite{Barabasi16}.

\begin{table}[tbp]
    \caption{Number of nodes and links for firm- and establishment-level networks.}
    \label{tbl:networkcount}
    \centering
    \begin{tabular}{l|rrrrr}
    & & & \multicolumn{1}{c}{Average Links} & \multicolumn{1}{c}{Median Links} & \multicolumn{1}{c}{90th Percentile} \\
    & \multicolumn{1}{c}{Node} & \multicolumn{1}{c}{Link} & \multicolumn{1}{c}{per Node} & \multicolumn{1}{c}{per Node} & \multicolumn{1}{c}{of Links per Node} \\
    \hline
    Firm-level & 157,537 & 480,507 & 6.1 & 3 & 7 \\
    Establishment-level & 183,951 & 919,982 & 10.0 & 4 & 17 \\    
    \end{tabular}
\end{table}

\begin{figure}[htpb]
\centering
\includegraphics[width=.75\linewidth]{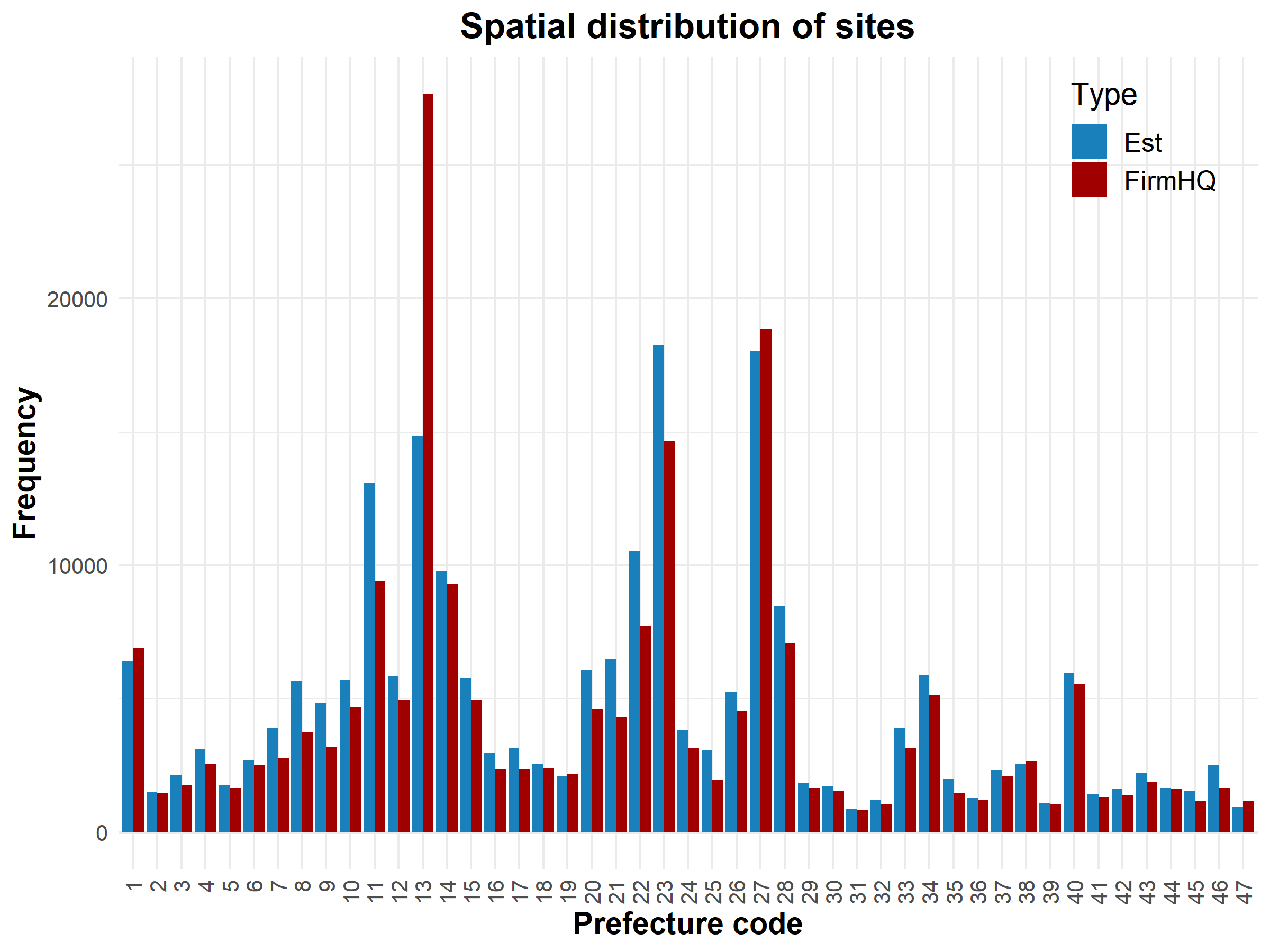}
\caption{Spatial distribution of firms and establishments: The red and blue bars show the distribution of firms and establishments, respectively, across prefectures. Prefecture codes are defined mostly from the north to the south. For example, Tokyo is code 13, whereas Saitama and Aichi are coded 11 and 23, respectively.}
\label{fig:space}
\end{figure}

Finally, the two networks differ markedly in their geographic distributions. Firm-level networks assign geographic location based on headquarters, whereas establishment-level networks capture the locations of individual production sites. Figure~\ref{fig:space} illustrates this distinction: Tokyo (Code 13) exhibits a high concentration of firm headquarters but fewer establishments, while industrial prefectures outside Tokyo such as Saitama (Code 11), Shizuoka (Code 22), and Aichi (Code 23) host many establishments whose firms are headquartered in Tokyo.

\vspace{10pt}
\vspace{10pt}

\section{Method}

We propose a parsimonious model in which a shock that externally imposes infeasibility of production on a particular firm or establishment propagates probabilistically through supply chains. Although economic agents typically optimize objectives such as profits or costs over the long run, our focus is on the short-run transmission of supply-side disruptions. Given this objective and the unique granularity of the data, a relatively simple modeling framework is desirable. More elaborate models may obscure whether observed propagation patterns arise from network structure or from strong behavioral assumptions embedded in the model, complicating comparisons across networks with different levels of granularity---particularly between firm-level networks and establishment--product-level networks.

A number of existing studies incorporate rich behavioral mechanisms to model shock propagation across economic entities. \cite{battiston_debtrank_2012,fujiwara_debtrank_2016,inoue_firm-level_2019,diem2022quantifying} A key distinction between these approaches and ours is that most existing models are deterministic, whereas our framework is probabilistic. Deterministic models offer transparent and interpretable outcomes but may understate systemic risk when propagation is inherently stochastic. To capture this uncertainty, we adopt a probabilistic approach inspired by Zhao et al.~\cite{zhao2019modelling}. Unlike their framework, which allows firms to search for substitute suppliers beyond their existing supply-chain partners, our model assumes a fixed set of suppliers and products determined by the observed establishment- and product-level network. Substitution in our model occurs only when alternative active suppliers already provide the same product, and substitution across different types of inputs is not allowed, implying zero elasticity of substitution and a fixed input structure. While restrictive, this assumption is consistent with the short-term nature of the disruptions we study and allows us to isolate the role of network granularity and product specificity.

\begin{figure}[htpb]
\centering
\includegraphics[width=0.8\linewidth]{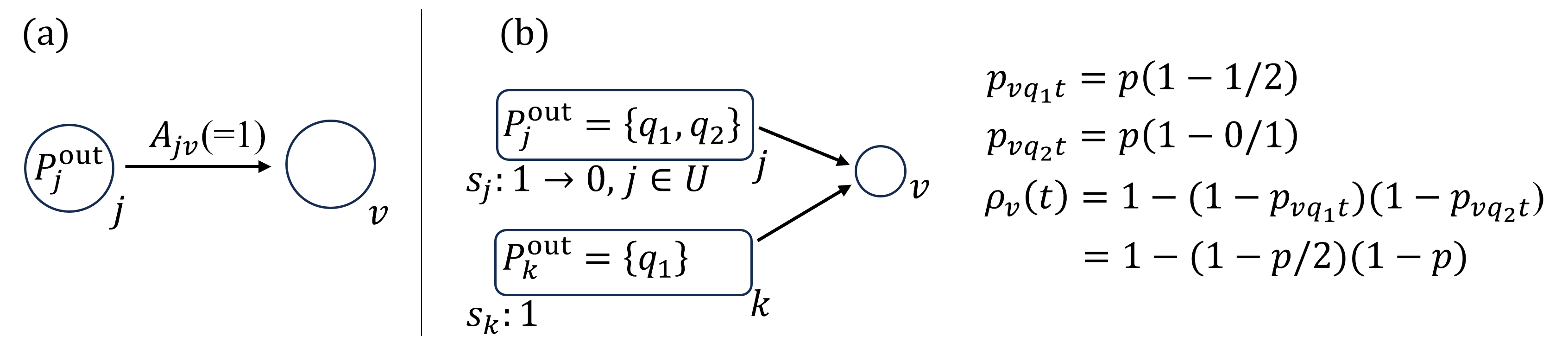}
\caption{Illustration of the propagation model: Panel (a) shows important notations in a supply chain flow from node $j$ to $v$. Panel (b) indicates an example of how the probability that a client $v$ becomes inactive, or $\rho_v$, is calculated, assuming that $v$ has two suppliers, $k$ producing product $q_1$ and $j$ producing products $q_1$ and $q_2$. 
}
\label{fig:model}
\end{figure}

We now formally describe the model. Figure~\ref{fig:model} provides a schematic overview and an illustrative example. Let $A_{ij}$ denote the adjacency matrix of the production network, where $i$ and $j$ represent entities (firms or establishments). $A_{ij}=1$ if entity $i$ supplies entity $j$, and $A_{ij}=0$ otherwise. Because the network is directed, $A_{ij}$ is generally asymmetric. We abstract from transaction volumes and focus solely on network topology. The set of output products produced by entity $i$ is denoted by $P^{\text{\scriptsize out}}_{i}$, which may contain one or multiple products.

\vspace{10pt}
\vspace{10pt}

The set of input products required by entity $i$ is derived from its suppliers as
\begin{equation}
P^{\text{\scriptsize in}}_{i} = \bigcup_{j} A_{ji} \otimes P^{\text{\scriptsize out}}_{j},
\end{equation}
where the operator $\otimes$ indicates that the products of supplier $j$ are included if and only if $A_{ji}=1$.

Each entity has a binary state $s_i(t)$ at time $t$, where $s_i(t)=1$ indicates that the entity is active and $s_i(t)=0$ indicates that it is inactive. Once an entity becomes inactive, it remains inactive for the remainder of the simulation, meaning that our simulation does not assume any recovery from supply chain disruptions. Let $U$ denote the set of entities that become newly inactive in the previous period $t-1$. Let $V$ be the set of active entities that have at least one supplier in $U$. For each entity $v \in V$, the probability that it becomes inactive at time $t$, denoted by $\rho_v(t)$, is given by
\begin{equation}
\rho_v(t) = 1 - \prod_{q \in P^{\text{\scriptsize in}}_{v}} (1 - p_{vqt}),
\label{eq:pro}
\end{equation}
where $p_{vqt}$ denotes the probability that a disruption propagates to entity $v$ through input product $q$ at time $t$. 

The product-specific propagation probability $p_{vqt}$ is defined as
\begin{equation}
p_{vqt} = p \left(1 - \frac{\sum_j s_j(t) A_{jv} \mathbf{1}_{P^{\text{\scriptsize out}}_{j}}(q)}{\sum_j A_{jv} \mathbf{1}_{P^{\text{\scriptsize out}}_{j}}(q)}\right),
\label{eq:compp}
\end{equation}
where $p$ is a baseline propagation probability parameter common to all entities and periods, and $\mathbf{1}_{P^{\text{\scriptsize out}}_{j}}(q)$ is an indicator function equal to one if supplier $j$ produces product $q$ and zero otherwise:
\begin{equation}
\mathbf{1}_{A}(x) =
\begin{cases}
1 & \text{if } x \in A, \\
0 & \text{if } x \notin A.
\end{cases}
\end{equation}
Equation~(\ref{eq:compp}) implies that propagation through product $q$ is mitigated in proportion to the share of suppliers of $q$ that remain active. When all suppliers of a required product are inactive, propagation occurs with probability $p$; when some suppliers remain active, the probability is proportionally reduced.

To illustrate, consider a client entity $v$ connected to two suppliers, as described in Panel (b) of Figure \ref{fig:model}: entity $j$, which produces products $\{q_1,q_2\}$, and entity $k$, which produces $\{q_1\}$. If entity $j$ becomes newly inactive at time $t$, propagation probabilities are evaluated for both $q_1$ and $q_2$. Because entity $k$ remains active and supplies $q_1$, the propagation probability through $q_1$ is mitigated. In contrast, no such mitigation applies to $q_2$, which is supplied only by entity $j$. For example, if the baseline probability $p$ is 0.5, then $\rho_v(t)$ equals 0.625. By contrast, in the absence of a substitution mechanism as specified in Equation~(\ref{eq:compp}), that is, when $p_{vqt}=0.5$ for both $q_1$ and $q_2$, $\rho_v(t)$ equals 0.75.

At the initial period, all entities are active. One entity is then randomly selected to become inactive, either from the full set of entities or from a restricted subset in simulations involving region-specific shocks. Propagation proceeds iteratively according to the rules described above until no new inactive entities emerge.  For each setting, we conduct 100{,}000 simulation runs and compute the share of inactive entities at the end of each run to measure the size of the propagation effect through supply chains. The baseline propagation probability $p$ is set at 0.5, because setting $p$ at any value from 0.2 to 1.0 provides qualitatively similar results although a larger $p$ results in a larger share of inactive entities. 

Finally, we note that richer formulations are possible. For example, one could model propagation at the level of individual outputs, conditioning both disruption and substitution on product-specific recipes. Such an extension would allow only those supplier outputs explicitly required for a given production process to affect propagation. While potentially more realistic, incorporating this level of detail would substantially increase computational complexity and obscure the comparison across alternative network constructions. Given our focus on isolating the effects of network and input granularity, we leave such extensions for future research.

\section{Simulation Results and Discussion}

Using the data and propagation model described above, we analyze how differences in the granularity of network nodes (firms versus establishments) and in the definition of inputs determining the degree of substitutability (industries versus products) affect the propagation of shocks through supply chains. To this end, we conduct simulations under five alternative settings, summarized in Table \ref{tbl:cases}.

\begin{figure}[tbp]
\centering
\includegraphics[width=.75\linewidth]{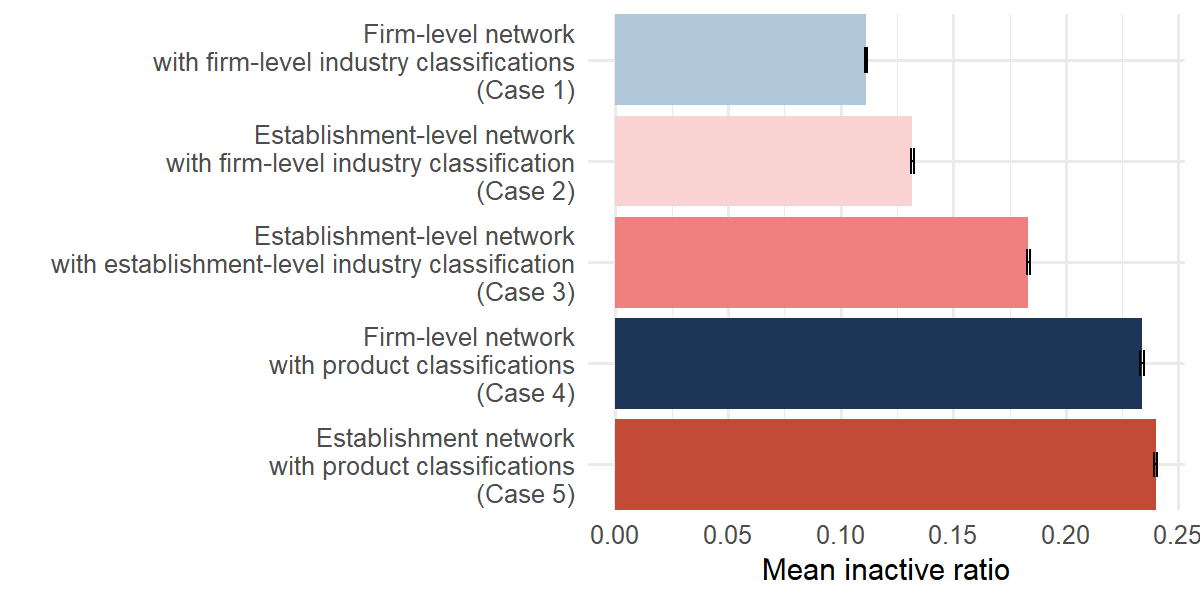}
\caption{Comparison of propagation effects under different granularity of nodes and input classifications:
Each bar reports the mean share of inactive entities (firms or establishments) for the following setting: (Case 1) firm-level network with industry-based inputs; (Case 2) establishment-level network with firm-level industry inputs; (Case 3) establishment-level network with establishment-level industry inputs; (Case 4) firm-level network with product-based inputs; and (Case 5) establishment-level network with product-based inputs. Error bars indicate standard errors. The propagation probability is set to $p=0.5$.}
\label{fig:estfirm}
\end{figure}

In case (1), we employ the firm-level supply-chain network in which inputs are classified by the primary industry of each firm as reported in the TSR data, following the existing firm-level supply-chain literature \cite{poledna2018does,inoue_firm-level_2019,diem2022quantifying,pichler_building_2023}. Case (2) also uses industry-based input classifications at the firm level but replaces firms with establishments as network nodes, thereby highlighting the effect of node granularity. Case (3) further refines the input definition by classifying inputs according to the industry of each establishment, as reported in the Census data, allowing industries to differ across establishments belonging to the same firm. In case (4), we revert to the firm-level network used in case (1) but classifies inputs using detailed product information aggregated from the firm’s establishments in the Census data. Finally, case (5) combines establishment-level nodes with product-level input classifications. Comparing cases (4) and (5) allows us to assess the role of node granularity while holding the granularity of inputs constant. Figure \ref{fig:estfirm} reports the average share of inactive nodes at the end of the simulations for each of the five cases.

\vspace{10pt}
\begin{table}[tbp]
\caption{Five cases in the simulation analysis}
\label{tbl:cases}
\centering
\begin{tabular}{l|c|c|c}
\hline
& \multicolumn{3}{c}{Definition of inputs} \\
\cline{2-4}
Network 
& Industry (firm level) 
& Industry (establishment level)
& Product \\ 
\hline
Firm-level           & (1) &       & (4) \\ 
Establishment-level  & (2) & (3)   & (5) \\ 
\hline
\end{tabular}
\begin{flushleft}
\footnotesize
Notes: This table shows the 5 simulation cases, which differ in the granularity of network nodes and the classification of inputs.
\end{flushleft}
\end{table}
\vspace{10pt}

\subsection{The role of node granularity in propagation}

We first examine how the granularity of network nodes affects the diffusion of economic shocks through supply chains. Specifically, we compare the firm-level and establishment-level networks under the same industry-based input classification at the firm level, corresponding to cases (1) and (2) in Table \ref{tbl:cases}. As shown by bars (1) and (2) in Figure \ref{fig:estfirm}, the mean inactive share is approximately 11\% in the firm-level network and increases to about 13\% in the establishment-level network. The difference is statistically significant. 

A similar comparison can be made between cases (4) and (5), which use product-level input definitions for firm-level and establishment-level networks, respectively. In this case as well, the establishment-level network exhibits a larger propagation effect than the firm-level network, as indicated by bars (4) and (5) in Figure \ref{fig:estfirm}.

Taken together, these results suggest that shocks tend to propagate more strongly in supply chains with finer node granularity. This finding is consistent with earlier studies showing that firm-level networks generate larger propagation effects than industry-level input--output analyses \cite{inoue2019firm, diem_estimating_2024}. However, whether establishment-level networks should exhibit stronger propagation than firm-level networks is theoretically ambiguous, due to two countervailing mechanisms.

On the one hand, the establishment-level network features a substantially higher average number of suppliers per node than the firm-level network (10.0 versus 6.1), while the total number of nodes differs only modestly (183{,}951 versus 157{,}537; see Table \ref{tbl:networkcount}). This higher connectivity increases small-world characteristics, which tend to amplify shock propagation \cite{Watts98}. On the other hand, a larger number of suppliers may also facilitate input substitution, thereby mitigating propagation. This mitigating effect is particularly relevant when suppliers tend to produce the identical inputs defined at the industry level, as in bars (1) and (2), or at the product level, as in bars (4) and (5).

Our finding that propagation is stronger in establishment-level networks indicates that the amplification effect arising from increased small-worldness dominates the substitution effect associated with higher supplier redundancy. At the same time, the magnitude of the difference in the propagation effect between firm- and establishment-level networks remains modest, as evidenced by the relatively small gaps between bars (1) and (2) and between bars (4) and (5). This pattern is consistent with the presence of these two opposing forces. This conclusion regarding node granularity remains unchanged when varying the baseline probability $p$ (SI Figure 2).

\subsection{The role of input granularity in propagation}

We next assess the importance of input granularity by comparing simulations that use the identical network structure but different input classifications. Specifically, we contrast industry-based and product-based input definitions within the firm-level network (bars (1) and (4)) and within the establishment-level network (bars (2), (3), and (5)) in Figure \ref{fig:estfirm}. Across both network types, defining inputs at the product level leads to substantially stronger propagation effects than defining them at the industry level.

This result suggests that industry classifications are too coarse to accurately capture substitutability among inputs and suppliers. However, as in the case of node granularity, there are two potentially opposing effects of input granularity, leading to theoretical ambiguity in its total effect, primarily because each entity (firm or establishment) is assigned a single industry classification, whereas an entity may produce multiple distinct products. 

On the one hand, entities operating within the same industry may produce distinct products that are not close substitutes, because the number of industries (595 as shown in Section 2.4) is substantially smaller than the number of products (2,396). Consequently, industry-based classifications overstate substitution possibilities and thus understate the extent of shock propagation. On the other hand, allowing entities to be associated with multiple product classifications can increase substitutability, because products of entities in different industries may overlap particularly when the entities' primary industry cannot fully specify their wide-variety products. Ex ante, therefore, it is not obvious whether product-based input definitions should generate stronger or weaker propagation than industry-based definitions.




Our results show that propagation is weaker under industry-based input definitions than under product-based definitions even at the firm level. This finding implies that the reduction in substitutability induced by finer input classifications dominates the potential increase in substitutability arising from entities producing multiple products. Consequently, product-based input models exhibit stronger propagation effects despite allowing for multiple classifications per entity.

Moreover, the quantitative impact of input granularity on propagation is considerably larger than that of node granularity. For example, employing the firm-level network, the mean inactive share increases from approximately 11\% under industry-based input definitions to about 23\% under product-based definitions (bars (1) and (4) in Figure \ref{fig:estfirm}). By contrast, holding the input definition fixed at the product level, the difference between the firm-level and establishment-level networks is much smaller (bars (4) and (5)). These findings indicate that the extent to which inputs are substitutable plays a more critical role in shaping propagation outcomes than the granularity of network nodes. Varying the baseline probability 
$p$ does not alter the qualitative role of input granularity (SI Figure~2).

\subsection{The role of establishment-level geographic information}

Finally, we investigate the importance of establishment-level geographic information when shocks originate in a specific region, such as those affected by natural disasters. The spatial distribution of firms differs markedly from that of establishments (Figure~\ref{fig:space}): corporate headquarters are highly concentrated in Tokyo, whereas establishments are geographically distributed across a wide range of regions.

A shock that directly affects a larger number of entities—for example, one that simultaneously impacts all firms headquartered in Tokyo—would mechanically generate larger aggregate effects. However, such scale effects are not the focus of this analysis. Instead, our objective is to clarify how the propagation of an initial shock of identical magnitude depends on the geographic representation of economic entities. Accordingly, we consider shocks applied to a single entity (a firm or an establishment), rather than to an entire region.

This distinction is crucial because aggregating firms geographically at the headquarters level can fundamentally distort the structure of supply-chain linkages. When firms are represented only by their headquarters locations, supply-chain links between establishments that span multiple regions may be artificially collapsed into a single location. As a result, propagation effects associated with geographic concentration can be overstated. By contrast, establishment-level geographic information allows us to more accurately capture how shocks propagate through spatially distributed production networks.

To examine how the aggregate impact of region-specific shocks varies with the spatial origin of the initial disruption, we modify the simulation by randomly selecting the initially inactive node from a given region rather than from the entire country. We focus on a comparison between shocks originating inside and outside Tokyo, given Tokyo’s distinctive role in the Japanese economy. To highlight the value of establishment-level geographic detail, we compare results obtained using firm-level and establishment-level networks. In these simulations, inputs are classified at the product level, corresponding to cases (4) and (5) in Table \ref{tbl:cases}, since the previous subsection demonstrated that industry-based classifications underestimate propagation effects.

\begin{figure}[tbp]
\centering
\includegraphics[width=.6\linewidth]{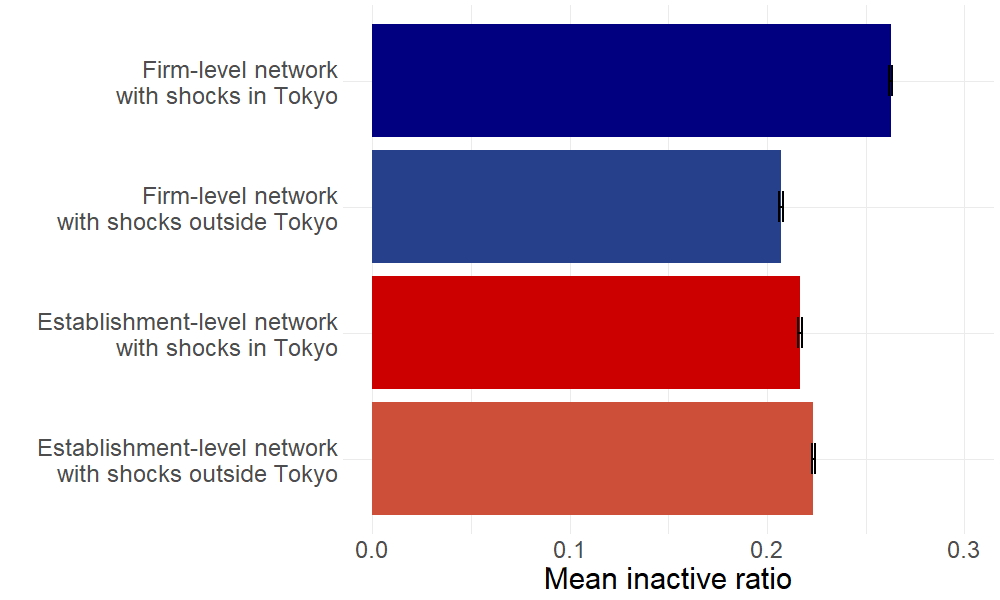}
\caption{Comparison of propagation effects for shocks originating inside and outside Tokyo. Each bar shows the mean share of inactive entities under the following settings: (1) firm-level network with shocks in Tokyo; (2) firm-level network with shocks outside Tokyo; (3) establishment-level network with shocks in Tokyo; and (4) establishment-level network with shocks outside Tokyo. Inputs are classified by product. Error bars indicate standard errors. The propagation probability is set to $p=0.5$.}
\label{fig:restrict}
\end{figure}

Figure \ref{fig:restrict} presents the results. In the firm-level network, which incorporates only headquarters locations, shocks originating in Tokyo generate substantially larger aggregate effects than shocks originating elsewhere, as shown by the difference between the first and second bars. In contrast, when using the establishment-level network, which assigns geographic locations to individual establishments, the effects of shocks inside and outside Tokyo are similar, as indicated by the bottom two bars.

This discrepancy reflects differences in network structure that are obscured in firm-level data. In the firm-level network, the average out-degree, or the average number of clients per node, is 7.24 for firms headquartered in Tokyo and 3.66 for firms headquartered elsewhere. In the establishment-level network, however, the corresponding figures are 5.61 for establishments in Tokyo and 7.79 for establishments outside Tokyo. Although average degree measures do not fully capture network complexity, these patterns suggest that major supply-chain hub establishments outside Tokyo are underrepresented in the firm-level network. Consequently, firm-level data tend to overstate the importance of Tokyo-based shocks while understating the impact of shocks originating in other regions.

As in the comparisons of network granularity and input classifications conducted in the previous sections, we vary the baseline probability $p$ to conduct robustness checks (SI Figure~3). We find that changes in $p$ do not alter the stylized facts described above.

Overall, these findings underscore the advantages of establishment-level supply-chain data for analyzing region-specific shocks, particularly when geographic concentration and network hubs play a central role in shock propagation.

\section{Conclusion}

This paper examined how the granularity of supply-chain data affects the propagation of economic shocks through production networks. Using large-scale simulations based on firm- and establishment-level supply-chain networks and alternative classifications of inputs, we showed that both the granularity of network nodes and input classifications play important roles in shaping propagation dynamics.

Our results highlight three main findings. First, shocks tend to propagate more strongly in establishment-level networks than in firm-level networks, reflecting the greater connectivity and small-world characteristics of more granular supply chains. However, the magnitude of this difference is relatively modest, suggesting that finer node granularity simultaneously amplifies propagation through increased connectivity while mitigating it through greater opportunities for supplier substitution. 

Second, the granularity of input classifications is quantitatively more important than the granularity of network nodes. Defining inputs at the product level substantially increases estimated propagation effects compared with industry-based classifications. This result reflects the dominance of finer input distinctions over the potential increase in substitutability arising from multiple product outputs. As a result, industry-based classifications tend to overstate substitutability and underestimate systemic vulnerability in supply chains. 

Third, incorporating establishment-level geographic information is crucial for assessing the impact of region-specific shocks of the same magnitude. Firm-level networks that assign locations only at the headquarters level tend to overstate the aggregate impact of shocks originating in major metropolitan areas, such as Tokyo, while understating the potential effects of shocks arising in other regions. By contrast, establishment-level data more accurately reflect the spatial distribution of production and the location of key network hubs, yielding a more balanced assessment of the geographic origins of systemic risk.

Taken together, these findings underscore the importance of detailed supply-chain data for accurately assessing systemic risk and economic resilience. Analyses based on coarse industry classifications or firm-level locations may substantially mischaracterize both the magnitude and the spatial distribution of propagation effects. From a policy perspective, our results suggest that efforts to enhance supply-chain resilience should be informed by granular information on production activities, input substitutability, and the geographic distribution of establishments. More broadly, this study highlights the value of establishment-level supply-chain networks as a foundation for evaluating the economic consequences of disruptions arising from natural disasters, geopolitical shocks, and other systemic risks.

Several directions for future research naturally follow from this study. First, the scope of the network can be expanded beyond manufacturing. In this paper, we restrict our analysis to manufacturing firms and establishments, using 157{,}537 out of 1{,}520{,}605 firms and 183{,}951 out of 5{,}156{,}063 establishments. Because output information is also available for other sectors, the framework developed in this paper can be extended to construct broader production networks. One potential challenge arises in the wholesale sector, where establishments primarily distribute products rather than transform them. Nevertheless, once production recipes are identified for non-wholesale sectors, wholesalers can be incorporated as intermediaries linking upstream suppliers and downstream clients. We leave the construction and analysis of such extended networks for future work.

A second avenue for future research is to examine alternative network structures generated by different link-construction algorithms. As discussed above, the algorithm employed in this study infers establishment-level links in a parsimonious manner, yielding the minimum number of connections consistent with observed transactions. However, additional information could be incorporated to refine link assignment. For example, prior studies show that geographic proximity increases the likelihood of inter-establishment transactions \cite{nakajima2012measuring}. Other potentially relevant factors include establishment size, sales volumes, shared financial institutions, or cross-shareholding relationships between firms \cite{krichene_emergence_2019}. Systematically comparing propagation risks across networks constructed using different algorithms and information sets represents an important direction for future research.

Finally, if the underlying production recipe were known, the filtering procedure proposed in this paper would be unnecessary, since supply-chain links could be constructed directly from the observed recipe. Even within our data, there may be limited cases in which recipe identification can be improved. In particular, when both a supplier and a client operate only a single establishment, their transaction relationship may provide a relatively clear indication of the underlying production recipe for the products involved. Exploiting such cases to refine inferred networks and improve the accuracy of propagation analysis is left for future work.

\section*{Acknowledgements}

This research was conducted as part of a project entitled ``Research on Relationships between Economic Networks and National Security'' undertaken at the Research Institute of Economy, Trade, and Industry (RIETI). The authors thank the seminar participants at RIETI.

\bibliography{reference}

\begin{thebibliography}{10}
\urlstyle{rm}
\expandafter\ifx\csname url\endcsname\relax
  \def\url#1{\texttt{#1}}\fi
\expandafter\ifx\csname urlprefix\endcsname\relax\def\urlprefix{URL }\fi
\expandafter\ifx\csname doiprefix\endcsname\relax\def\doiprefix{DOI: }\fi
\providecommand{\bibinfo}[2]{#2}
\providecommand{\eprint}[2][]{\url{#2}}

\bibitem{Carvalho2016}
\bibinfo{author}{Carvalho, V.~M.}, \bibinfo{author}{Nirei, M.}, \bibinfo{author}{Saito, Y.~U.} \& \bibinfo{author}{Tahbaz-Salehi, A.}
\newblock \bibinfo{title}{Supply chain disruptions: Evidence from the great east japan earthquake}.
\newblock \bibinfo{type}{Tech. Rep.} \bibinfo{number}{No. 17-5}, \bibinfo{institution}{Columbia Business School Research Paper} (\bibinfo{year}{2016}).

\bibitem{Barrot2016}
\bibinfo{author}{Barrot, J.-N.} \& \bibinfo{author}{Sauvagnat, J.}
\newblock \bibinfo{journal}{\bibinfo{title}{Input specificity and the propagation of idiosyncratic shocks in production networks}}.
\newblock {\emph{\JournalTitle{The Quarterly Journal of Economics}}} \textbf{\bibinfo{volume}{131}}, \bibinfo{pages}{1543--1592} (\bibinfo{year}{2016}).

\bibitem{tierney1997business}
\bibinfo{author}{Tierney, K.~J.}
\newblock \bibinfo{journal}{\bibinfo{title}{Business impacts of the northridge earthquake}}.
\newblock {\emph{\JournalTitle{Journal of Contingencies and crisis management}}} \textbf{\bibinfo{volume}{5}}, \bibinfo{pages}{87--97} (\bibinfo{year}{1997}).

\bibitem{pelling2002macro}
\bibinfo{author}{Pelling, M.}, \bibinfo{author}{{\"O}zerdem, A.} \& \bibinfo{author}{Barakat, S.}
\newblock \bibinfo{journal}{\bibinfo{title}{The macro-economic impact of disasters}}.
\newblock {\emph{\JournalTitle{Progress in Development Studies}}} \textbf{\bibinfo{volume}{2}}, \bibinfo{pages}{283--305} (\bibinfo{year}{2002}).

\bibitem{bak1993aggregate}
\bibinfo{author}{Bak, P.}, \bibinfo{author}{Chen, K.}, \bibinfo{author}{Scheinkman, J.} \& \bibinfo{author}{Woodford, M.}
\newblock \bibinfo{journal}{\bibinfo{title}{Aggregate fluctuations from independent sectoral shocks: self-organized criticality in a model of production and inventory dynamics}}.
\newblock {\emph{\JournalTitle{Ricerche economiche}}} \textbf{\bibinfo{volume}{47}}, \bibinfo{pages}{3--30} (\bibinfo{year}{1993}).

\bibitem{gatti2005new}
\bibinfo{author}{Gatti, D.~D.} \emph{et~al.}
\newblock \bibinfo{journal}{\bibinfo{title}{A new approach to business fluctuations: heterogeneous interacting agents, scaling laws and financial fragility}}.
\newblock {\emph{\JournalTitle{Journal of Economic behavior \& organization}}} \textbf{\bibinfo{volume}{56}}, \bibinfo{pages}{489--512} (\bibinfo{year}{2005}).

\bibitem{acemoglu2012network}
\bibinfo{author}{Acemoglu, D.}, \bibinfo{author}{Carvalho, V.~M.}, \bibinfo{author}{Ozdaglar, A.} \& \bibinfo{author}{Tahbaz-Salehi, A.}
\newblock \bibinfo{journal}{\bibinfo{title}{The network origins of aggregate fluctuations}}.
\newblock {\emph{\JournalTitle{Econometrica}}} \textbf{\bibinfo{volume}{80}}, \bibinfo{pages}{1977--2016} (\bibinfo{year}{2012}).

\bibitem{haimes2001leontief}
\bibinfo{author}{Haimes, Y.~Y.} \& \bibinfo{author}{Jiang, P.}
\newblock \bibinfo{journal}{\bibinfo{title}{Leontief-based model of risk in complex interconnected infrastructures}}.
\newblock {\emph{\JournalTitle{Journal of Infrastructure systems}}} \textbf{\bibinfo{volume}{7}}, \bibinfo{pages}{1--12} (\bibinfo{year}{2001}).

\bibitem{okuyama2004measuring}
\bibinfo{author}{Okuyama, Y.}, \bibinfo{author}{Hewings, G.~J.} \& \bibinfo{author}{Sonis, M.}
\newblock \bibinfo{title}{Measuring economic impacts of disasters: interregional input-output analysis using sequential interindustry model}.
\newblock In \emph{\bibinfo{booktitle}{Modeling spatial and economic impacts of disasters}}, \bibinfo{pages}{77--101} (\bibinfo{publisher}{Springer}, \bibinfo{year}{2004}).

\bibitem{inoue_firm-level_2019}
\bibinfo{author}{Inoue, H.} \& \bibinfo{author}{Todo, Y.}
\newblock \bibinfo{journal}{\bibinfo{title}{Firm-level propagation of shocks through supply-chain networks}}.
\newblock {\emph{\JournalTitle{Nature Sustainability}}} \textbf{\bibinfo{volume}{2}}, \bibinfo{pages}{841--847}, \doiprefix\url{10.1038/s41893-019-0351-x} (\bibinfo{year}{2019}).
\newblock \bibinfo{note}{Number: 9 Publisher: Nature Publishing Group}.

\bibitem{Bossut2024granular}
\bibinfo{author}{Bossut, M.} \emph{et~al.}
\newblock \bibinfo{title}{A call for granular supply network data for navigating the climate transition}.
\newblock \bibinfo{type}{T20 Policy Brief} \bibinfo{number}{5166428}, \bibinfo{institution}{SSRN} (\bibinfo{year}{2024}).

\bibitem{diem2024estimating}
\bibinfo{author}{Diem, C.}, \bibinfo{author}{Borsos, A.}, \bibinfo{author}{Reisch, T.}, \bibinfo{author}{Kert{\'e}sz, J.} \& \bibinfo{author}{Thurner, S.}
\newblock \bibinfo{journal}{\bibinfo{title}{Estimating the loss of economic predictability from aggregating firm-level production networks}}.
\newblock {\emph{\JournalTitle{PNAS nexus}}} \textbf{\bibinfo{volume}{3}}, \bibinfo{pages}{pgae064} (\bibinfo{year}{2024}).

\bibitem{poledna2018does}
\bibinfo{author}{Poledna, S.} \emph{et~al.}
\newblock \bibinfo{journal}{\bibinfo{title}{When does a disaster become a systemic event? estimating indirect economic losses from natural disasters}}.
\newblock {\emph{\JournalTitle{arXiv preprint arXiv:1801.09740}}}  (\bibinfo{year}{2018}).

\bibitem{diem2022quantifying}
\bibinfo{author}{Diem, C.}, \bibinfo{author}{Borsos, A.}, \bibinfo{author}{Reisch, T.}, \bibinfo{author}{Kert{\'e}sz, J.} \& \bibinfo{author}{Thurner, S.}
\newblock \bibinfo{journal}{\bibinfo{title}{Quantifying firm-level economic systemic risk from nation-wide supply networks}}.
\newblock {\emph{\JournalTitle{Scientific reports}}} \textbf{\bibinfo{volume}{12}}, \bibinfo{pages}{7719} (\bibinfo{year}{2022}).

\bibitem{pichler_building_2023}
\bibinfo{author}{Pichler, A.} \emph{et~al.}
\newblock \bibinfo{journal}{\bibinfo{title}{Building an alliance to map global supply networks}}.
\newblock {\emph{\JournalTitle{Science}}} \textbf{\bibinfo{volume}{382}}, \bibinfo{pages}{270--272}, \doiprefix\url{10.1126/science.adi7521} (\bibinfo{year}{2023}).
\newblock \bibinfo{note}{Publisher: American Association for the Advancement of Science}.

\bibitem{inoue2019firm}
\bibinfo{author}{Inoue, H.} \& \bibinfo{author}{Todo, Y.}
\newblock \bibinfo{journal}{\bibinfo{title}{Firm-level propagation of shocks through supply-chain networks}}.
\newblock {\emph{\JournalTitle{Nature Sustainability}}} \textbf{\bibinfo{volume}{2}}, \bibinfo{pages}{841--847} (\bibinfo{year}{2019}).

\bibitem{inoue2023trade}
\bibinfo{author}{Inoue, H.} \& \bibinfo{author}{Todo, Y.}
\newblock \bibinfo{journal}{\bibinfo{title}{Disruption of international trade and its propagation through firm-level domestic supply chains: A case of japan}}.
\newblock {\emph{\JournalTitle{PLOS ONE}}} \textbf{\bibinfo{volume}{18}}, \bibinfo{pages}{e0294574} (\bibinfo{year}{2023}).

\bibitem{Berthou2024granular}
\bibinfo{author}{Berthou, A.}, \bibinfo{author}{Haramboure, A.} \& \bibinfo{author}{Samek, L.}
\newblock \bibinfo{title}{Mapping and testing product-level vulnerabilities in granular production networks}.
\newblock \bibinfo{type}{OECD Science, Technology and Industry Working Papers} \bibinfo{number}{2024/02}, \bibinfo{institution}{OECD} (\bibinfo{year}{2024}).

\bibitem{SBJ2021}
\bibinfo{author}{{Statistics Bureau of Japan}}.
\newblock \bibinfo{title}{{Economic Census for Business Activity}}.
\newblock \bibinfo{howpublished}{\url{https://www.stat.go.jp/english/data/e-census/2021/index.html}} (\bibinfo{year}{2021}).
\newblock \bibinfo{note}{Accessed: 08.02.2024}.

\bibitem{JSIC2013}
\bibinfo{author}{{Ministry of Internal Affairs and Communications, Japan}}.
\newblock \bibinfo{title}{{Japan Standard Industrial Classification (Rev. 13, October 2013)}}.
\newblock \bibinfo{howpublished}{\url{https://www.soumu.go.jp/english/dgpp_ss/seido/sangyo/index13.htm }} (\bibinfo{year}{2013}).
\newblock \bibinfo{note}{Accessed: 08.16.2024}.

\bibitem{Barabasi16}
\bibinfo{author}{Barab{\'a}si, A.-L.}
\newblock \emph{\bibinfo{title}{Network science}} (\bibinfo{publisher}{Cambridge University Press}, \bibinfo{year}{2016}).

\bibitem{battiston_debtrank_2012}
\bibinfo{author}{Battiston, S.}, \bibinfo{author}{Puliga, M.}, \bibinfo{author}{Kaushik, R.}, \bibinfo{author}{Tasca, P.} \& \bibinfo{author}{Caldarelli, G.}
\newblock \bibinfo{journal}{\bibinfo{title}{Debtrank: Too central to fail? financial networks, the fed and systemic risk}}.
\newblock {\emph{\JournalTitle{Scientific reports}}} \textbf{\bibinfo{volume}{2}}, \bibinfo{pages}{1--6} (\bibinfo{year}{2012}).

\bibitem{fujiwara_debtrank_2016}
\bibinfo{author}{Fujiwara, Y.}, \bibinfo{author}{Terai, M.}, \bibinfo{author}{Fujita, Y.} \& \bibinfo{author}{Souma, W.}
\newblock \bibinfo{journal}{\bibinfo{title}{Debtrank analysis of financial distress propagation on a production network in japan}}.
\newblock {\emph{\JournalTitle{RIETI Discussion Paper Series}}} \textbf{\bibinfo{volume}{15}} (\bibinfo{year}{2016}).

\bibitem{zhao2019modelling}
\bibinfo{author}{Zhao, K.}, \bibinfo{author}{Zuo, Z.} \& \bibinfo{author}{Blackhurst, J.~V.}
\newblock \bibinfo{journal}{\bibinfo{title}{Modelling supply chain adaptation for disruptions: An empirically grounded complex adaptive systems approach}}.
\newblock {\emph{\JournalTitle{Journal of operations Management}}} \textbf{\bibinfo{volume}{65}}, \bibinfo{pages}{190--212} (\bibinfo{year}{2019}).

\bibitem{diem_estimating_2024}
\bibinfo{author}{Diem, C.}, \bibinfo{author}{Borsos, A.}, \bibinfo{author}{Reisch, T.}, \bibinfo{author}{Kert{\'e}sz, J.} \& \bibinfo{author}{Thurner, S.}
\newblock \bibinfo{journal}{\bibinfo{title}{Estimating the loss of economic predictability from aggregating firm-level production networks}}.
\newblock {\emph{\JournalTitle{PNAS nexus}}} \textbf{\bibinfo{volume}{3}}, \bibinfo{pages}{pgae064} (\bibinfo{year}{2024}).

\bibitem{Watts98}
\bibinfo{author}{Watts, D.~J.} \& \bibinfo{author}{Strogatz, S.~H.}
\newblock \bibinfo{journal}{\bibinfo{title}{Collective dynamics of `small-world' networks}}.
\newblock {\emph{\JournalTitle{Nature}}} \textbf{\bibinfo{volume}{393}}, \bibinfo{pages}{440--442} (\bibinfo{year}{1998}).

\bibitem{nakajima2012measuring}
\bibinfo{author}{Nakajima, K.}, \bibinfo{author}{Saito, Y.~U.} \& \bibinfo{author}{Uesugi, I.}
\newblock \bibinfo{journal}{\bibinfo{title}{Measuring economic localization: Evidence from japanese firm-level data}}.
\newblock {\emph{\JournalTitle{Journal of the Japanese and International Economies}}} \textbf{\bibinfo{volume}{26}}, \bibinfo{pages}{201--220} (\bibinfo{year}{2012}).

\bibitem{krichene_emergence_2019}
\bibinfo{author}{Krichene, H.} \emph{et~al.}
\newblock \bibinfo{journal}{\bibinfo{title}{The emergence of properties of the {Japanese} production network: {How} do listed firms choose their partners?}}
\newblock {\emph{\JournalTitle{Social Networks}}} \textbf{\bibinfo{volume}{59}}, \bibinfo{pages}{1--9}, \doiprefix\url{10.1016/j.socnet.2019.05.002} (\bibinfo{year}{2019}).

\end{thebibliography}

\section*{Author contributions statement}

H.I conceived the study. H.I. and Y.T conducted the analyses and wrote and reviewed the manuscript.

\section*{Data availability}
The data underlying this study can be obtained from Tokyo Shoko Research, Ltd., the Ministry of Internal Affairs and Communications of Japan, and the Ministry of Economy, Trade and Industry of Japan. However, access to these data is subject to licensing and confidentiality restrictions, and they are therefore not publicly available. The datasets used and/or analyzed in this study may be made available from the corresponding author upon reasonable request, subject to the approval of Tokyo Shoko Research, Ltd., the Ministry of Internal Affairs and Communications, the Ministry of Economy, Trade and Industry, and Research Institute of Economy, Trade and Industry.

\section*{Funding information}
This work was supported by JSPS KAKENHI Grant Numbers JP22K18533, JP23K20626, JP23K25520, JP25K01454, JST PRESTO Grant Number JPMJPR21R2, and the Asahi Glass Foundation.

\end{document}